%% file: main.tex
\PassOptionsToPackage{table}{xcolor}

\documentclass[a4paper,fleqn]{cas-dc}

\usepackage[authoryear]{natbib}
\usepackage{lmodern}
\usepackage{listings}
\usepackage{fontawesome5}
\usepackage{array}

\definecolor{umcblue}{RGB}{0,112,186}
\definecolor{umclight}{RGB}{232,244,252}
\definecolor{umcdark}{RGB}{0,45,98}

\def\tsc#1{\csdef{#1}{\textsc{\lowercase{#1}}\xspace}}
\tsc{WGM}
\tsc{QE}
\tsc{EP}
\tsc{PMS}
\tsc{BEC}
\tsc{DE}

\renewcommand{\printorcid}{}

\begin{document}
\hypersetup{bookmarksnumbered=true}
\def\floatpagepagefraction{1}
\def\textpagefraction{.001}

\shorttitle{Review of Prompt-Driven Target Volume Segmentation in Radiotherapy}

\shortauthors{Kombat et al.}

\title[mode=title]{From Vision to Context: A Review of Prompt-Driven Target Volume Segmentation in Radiotherapy}

\author[1]{Bakpen Kombat}
\cormark[1]
\ead{b.kombat@umcutrecht.nl}
\credit{Conceptualization, Screening, Formal Analysis, Original Draft, Editing}

\author[1]{Guus Grimbergen}
\credit{Conceptualization, Screening, Supervision, Review}

\author[2]{Maarten L. Terpstra}
\credit{Conceptualization, Supervision, Review}

\author[1]{Gert J. Meijer}
\credit{Conceptualization, Supervision, Review}

\author[1]{Lois A. Daamen}
\credit{Conceptualization, Supervision, Review}

\affiliation[1]{organization={Department of Radiation Oncology, University Medical Center Utrecht},
    city={Utrecht},
    country={The Netherlands}}

\affiliation[2]{
    organization={Computational Imaging Group for MR Diagnostics and Therapy, Center for Image Sciences, University Medical Center Utrecht},
    city={Utrecht},
    country={The Netherlands}
}

\cortext[cor1]{Corresponding author}

\begin{abstract}
Accurate target volume (TV) segmentation is essential for radiotherapy (RT) planning but remains time-consuming. Moreover, the final contour depends on the expertise of the radiation oncologist. Although deep learning has improved medical image segmentation, image-only models cannot encode the clinical reasoning that governs TV delineation, particularly for the clinical target volume (CTV), where boundaries are defined by clinical knowledge rather than imaging features alone. Prompt-driven segmentation has recently emerged as a strategy to address this limitation. This work aims to review these approaches for RT TV segmentation with four objectives: (1) compare the different prompting techniques used; (2) compare their performance against image-only baselines; (3) assess how they are evaluated against standard metrics, including the Dice similarity coefficient (DSC), 95th Hausdorff distance, sensitivity, precision, and clinical assessment; and (4) identify the gaps needed for clinical translation. We systematically searched PubMed, Embase, and arXiv for studies incorporating geometric, clinical text, or interactive prompts into deep learning models for RT TV delineation. Thirteen studies met the inclusion criteria, spanning head and neck, breast, prostate, lung, cervix, and rectal cancers. Seven studies used clinical text prompting, four used geometric prompting, and two used interactive prompting; three combined modalities. Reported DSC ranged from 0.67 to 0.93, compared to 0.65 to 0.91 for image-only baselines. None reported sensitivity or precision. Clinical text prompting appears well suited to CTV delineation. Evidence remains limited, and standardized prompt engineering and clinical validation are needed before these approaches can be translated into routine RT practice.
\end{abstract}



\maketitle

\input{sections/introduction}

\input{sections/methods}

\input{sections/results}

\input{sections/discussion}
\input{sections/conclusion}

\printcredits

\input{sections/declarations}

\bibliographystyle{cas-model2-names}
\bibliography{references}

\clearpage
\input{sections/appendix}

\end{document}

%% file: sections/introduction.tex
\section{Introduction}
\label{section 1}

Radiotherapy (RT) is a cornerstone of modern oncology and is estimated to be required in approximately half of all cancer patients worldwide~\citep{zhu2024global,siegel2024}. A critical component of effective RT planning is the accurate delineation of target volumes (TV), including the Gross Tumor Volume (GTV), Clinical Target Volume (CTV), and Internal Target Volume (ITV), as well as organs at risk (OARs), as defined by the International Commission on Radiation Units and Measurements (ICRU)~\citep{icru50,icru62,icru83}. These contours directly determine the delivered dose distribution, with the primary aim of delivering a high dose to the TV while sparing surrounding OARs. Errors introduced during delineation may therefore compromise tumor control, increase the risk of toxicity to surrounding tissues, and propagate through subsequent stages of treatment planning and delivery. 

Manual TV delineation remains one of the most time-consuming and error-prone steps in the RT workflow~\citep{darby2026}. This error-proneness is most evident in interobserver variability (IOV). A large body of evidence has shown that IOV in TV contouring is a major source of geometric uncertainty in RT planning, exceeding other well-established uncertainties such as patient positioning and organ motion~\citep{guzene2023,jamema2016}. Substantial IOV has been reported across multiple cancer sites, including lung, prostate, and esophageal cancer~\citep{guzene2023,jamema2016,vollenbrock2019}, with dosimetric analyzes further demonstrating that such variability results in clinically relevant deviations in dose delivery~\citep{chen2025,peng2018}. IOV arises from multiple factors, including radiation oncologist expertise, image quality, and the inherent subjectivity in interpreting ambiguous tumor boundaries on medical images~\citep{guzene2023}.

The clinical introduction of deep learning has reshaped contouring in RT. Deep learning has become the dominant approach for medical image segmentation because it can learn features directly from annotated data~\citep{litjens2017,shen2017}. Image-based deep learning architectures, such as U-Net~\citep{ronneberger2015} and its extensions, including 3D U-Net~\citep{cicek2016}, V-Net~\citep{milletari2016}, Attention U-Net~\citep{oktay2018}, and UNet++~\citep{zhou2018}, together with the self-configuring nnU-Net framework~\citep{isensee2021}, and more recently, transformer-based designs such as TransUNet~\citep{chen2021}, UNETR~\citep{hatamizadeh2022}, Swin-UNETR~\citep{liu2021swin,tang2022}, and nnFormer~\citep{zhou2023}, have progressively improved segmentation accuracy across anatomical structures and imaging modalities. In RT, deep learning-based OAR auto-segmentation has shown good agreement with expert contours across multiple anatomical sites~\citep{cardenas2019}. Multi-institutional studies have reported reductions in contouring time and improved consistency~\citep{lucido2025}, and commercial OAR auto-segmentation tools are now being integrated into clinical workflows in major RT centers~\citep{doolan2023}.

However, these advances have not translated to TV delineation to the same extent, and clinical adoption of deep learning models for GTV and CTV segmentation remains limited~\citep{debooij2024}. The difference arises from the nature of the two tasks. OARs such as the lungs, heart, and spinal cord occupy predictable anatomical locations with well-defined imaging boundaries, allowing models to learn their extent directly from annotated scans~\citep{cardenas2019,doolan2023}. However, TV delineation is clinically more complex, in particular for the CTV. For many cancer sites, the CTV boundary is not directly visible on imaging but is instead determined by tumor biology, patterns of microscopic spread, anatomical knowledge of which structures tumors can realistically invade, clinical staging, and institutional contouring guidelines~\citep{brierley2017,debooij2024,guzene2023}. As none of these features are encoded in the imaging, a segmentation model trained only on image data can learn only the relationship between image appearance and the contours provided during training, without access to the clinical information that determines the CTV extent. This limitation is not expected to be resolved by developing larger architectures or training on more image data alone.

This challenge has motivated the development of prompt-driven models (PDMs): segmentation models that incorporate information that cannot be derived from the image alongside the imaging data. Depending on the clinical context, this information, referred to as a prompt, may be geometric, textual, or interactive.  A major step in this direction was the Segment Anything Model (SAM)~\citep{kirillov2023}, a general-purpose vision foundation model capable of producing segmentation from simple geometric prompts such as points and bounding boxes. Its successor, SAM~2, further extended this capability to video segmentation~\citep{ravi2024}. However, without task-specific adaptation, SAM performs inconsistently across medical imaging applications because of the substantial domain gap between natural and medical images~\citep{ma2024,mazurowski2023}. Several medically adapted variants have been developed to overcome this gap, including MedSAM~\citep{ma2024}, SAM-Med2D~\citep{cheng2023}, SAM-Med3D~\citep{wang2023}, and nnInteractive~\citep{nninteractive2025}. 

Prompt learning methods have also shown that learnable prompts can efficiently adapt foundation models to downstream tasks without retraining the entire network~\citep{zhou2022coop}. Clinical text prompting has drawn on large language models (LLMs) such as BERT~\citep{devlin2019}, BioBERT~\citep{lee2020biobert}, BioGPT~\citep{luo2022}, and GPT-4~\citep{openai2023}, together with vision-language models, including CLIP~\citep{radford2021}, BLIP~\citep{li2022blip}, and LLaVA~\citep{liu2023llava}, to encode textual information from clinical reports and contouring guidelines alongside imaging features~\citep{oh2024,rajendran2025}.

Despite this growing body of work, PDMs for TV segmentation have not yet been systematically reviewed. Existing reviews have focused on OAR auto-segmentation~\citep{cardenas2019,lucido2025}, image-only CTV delineation~\citep{matoska2024}, the barriers to translating these methods to clinical practice ~\citep{debooij2024}, or SAM and its medical adaptations~\citep{huang2024}, without comparing the different prompting strategies or their application across cancer sites.  

This review addresses these gaps by providing a systematic review of PDMs for TV segmentation in RT. We compare prompting strategies and their performance against image-only baselines across multiple cancer sites, identify current methodological limitations, and discuss the remaining challenges for clinical translation.

%% file: sections/methods.tex
\section{Methods}
\label{section 2}
This review identified studies that developed prompt-driven deep learning approaches for TV segmentation in RT using a predefined search strategy and eligibility criteria. It was prospectively registered in PROSPERO (\href{https://www.crd.york.ac.uk/PROSPERO/view/CRD420261355774}{CRD420261355774}) and is reported in accordance with the PRISMA 2020 guidelines~\citep{page2021}.

\subsection{Prompt Modality Classification}
\label{sec:taxonomy}

\begin{figure*}[!t]
\centering
\setlength{\fboxsep}{12pt}%
\colorbox{white}{%
  \begin{minipage}{\dimexpr\textwidth-24pt}
    \centering\vspace{8pt}
    \setlength{\arrayrulewidth}{0.6pt}%
    \arrayrulecolor{gray!45}%
    \begin{tabular}{@{}c@{\hspace{8pt}}|@{\hspace{8pt}}c@{}}
      \includegraphics[width=7.0cm]{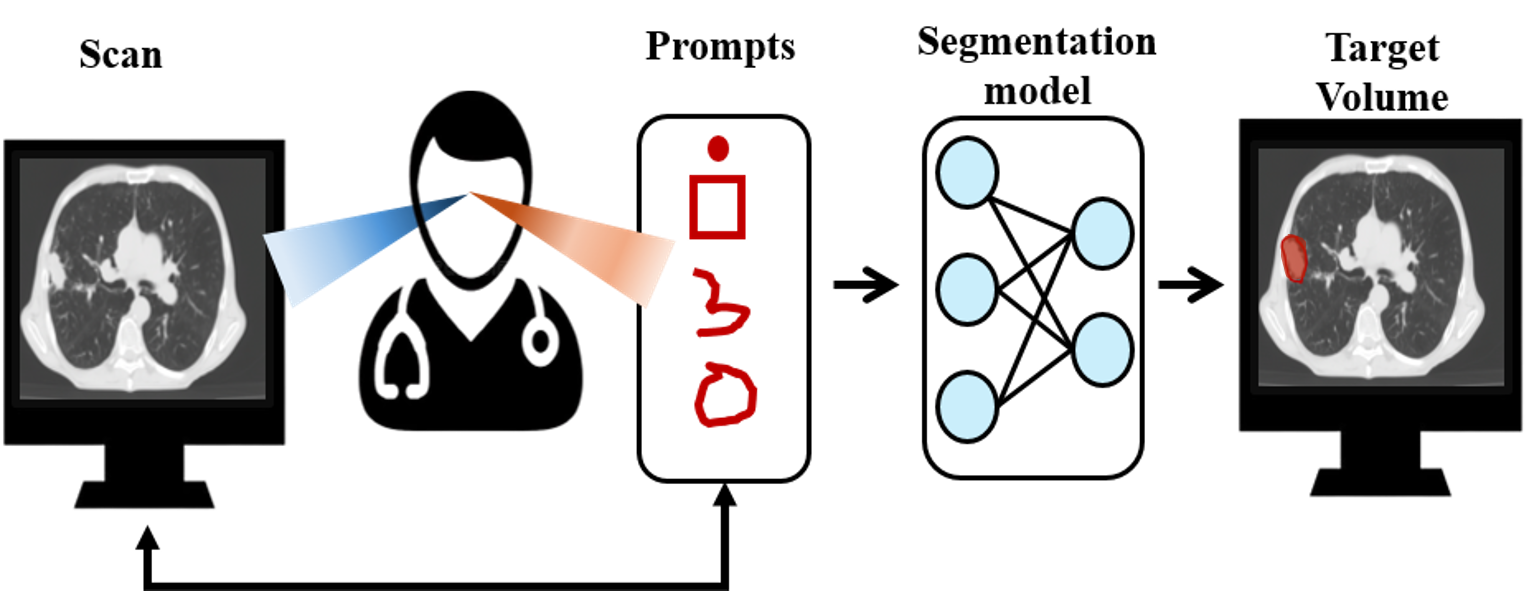} &
      \includegraphics[width=7.0cm]{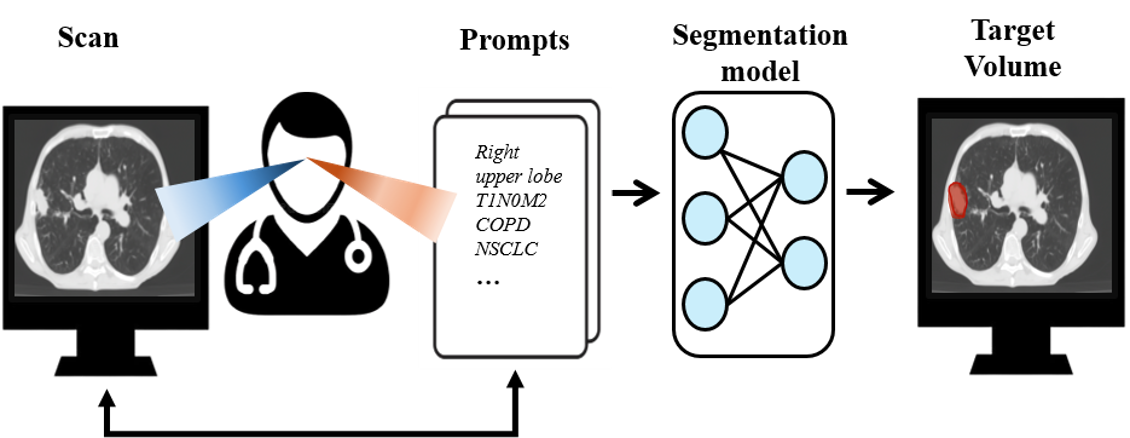} \\[4pt]
      \small\textbf{(a) Geometric prompting} &
      \small\textbf{(b) Clinical text prompting}
    \end{tabular}
    \arrayrulecolor{black}%
    \par\vspace{10pt}
    {\color{gray!45}\hrule height 0.6pt}%
    \vspace{10pt}
    \includegraphics[width=7.0cm]{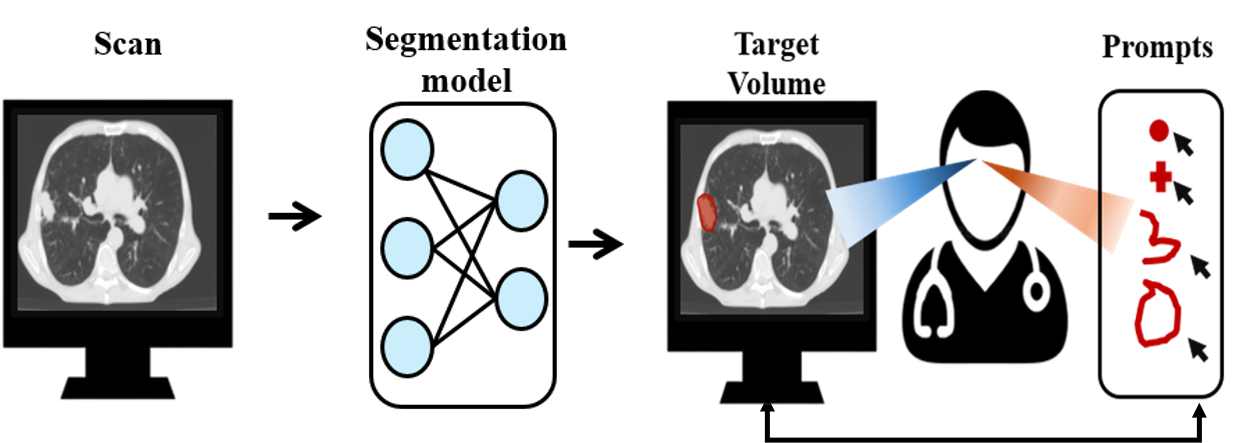}\\[4pt]
    \small\textbf{(c) Interactive prompting}
    \vspace{8pt}
  \end{minipage}%
}
\caption{The three prompt modalities reviewed in this paper. (a) Geometric prompting: spatial prompts such as points, bounding boxes, and binary masks. (b) Clinical text prompting: structured clinical information, including tumor stage, laterality, and contouring guidelines, is encoded as text and fused with imaging features. (c) Interactive prompting: the model generates an initial segmentation that the clinician reviews and refines by providing additional spatial feedback in a closed loop.}
\label{fig:taxonomy}
\end{figure*}

Studies were classified according to the type of prompt incorporated into the segmentation process. Three prompt modalities were identified (Figure~\ref{fig:taxonomy}): (1) geometric prompting, in which spatial inputs such as points, bounding boxes, or binary masks indicate the approximate location of the TV; (2) clinical text prompting, in which structured clinical data such as TV location and TNM staging, or unstructured information such as clinical reports and contouring guidelines, are encoded as text and fused with imaging features; and (3) interactive prompting, in which a clinician refines an initial model segmentation through corrective inputs. Studies employing more than one
modality were classified under all applicable categories.

\subsection{Search Strategy}
A literature search was conducted on April 8, 2026, to identify studies published between January 1, 2022, and April 8, 2026. This period was selected to capture the emergence of prompt-driven and multimodal AI approaches following the introduction of foundation models and LLMs. Three electronic databases were searched: PubMed, Embase, and arXiv; the full search strategies are provided in \hyperref[sec:search_strategies]{Appendix~A}.

\subsection{Eligibility Criteria}
\label{subsec:eligibility}

Studies were included if they:

\begin{itemize}
\item Employed a deep learning-based segmentation model.
\item Incorporated at least one prompting mechanism, such as geometric prompts
(e.g., points, bounding boxes, scribbles, or masks), clinical text inputs
(e.g., staging information, clinical reports, or structured clinical
data), or interactive/corrective prompting.
\item Reported quantitative segmentation performance metrics (e.g., Dice Similarity Coefficient (DSC), 95th-percentile Hausdorff Distance (HD95), sensitivity, or precision) or clinical evaluation by a radiation oncologist.
\end{itemize}

Studies were excluded if they:

\begin{itemize}
\item Focused exclusively on OAR segmentation without TV delineation.
\item Were review articles, meta-analyses, editorials, commentaries, letters, or conference abstracts.
\item Did not incorporate a prompting mechanism as defined above.
\item Employed non-deep learning methods, including atlas-based, traditional image processing, or deformable registration approaches without neural networks.
\item Reported only qualitative findings without quantitative segmentation performance metrics.
\item Were not published in English.
\end{itemize}

\subsection{Study Selection}

Records retrieved from the database searches were imported into Rayyan~\citep{rayyan} for screening. Titles and abstracts were screened independently by two reviewers (B.K. and G.G.) according to the predefined eligibility criteria. Disagreements were resolved through discussion until consensus was reached. Full-text assessment of potentially eligible studies was subsequently performed by B.K. Studies that met all inclusion criteria (Section~\ref{subsec:eligibility}) were included in the final review.

\subsection{Data Extraction and Synthesis}
\label{subsec:extraction}

For each included study, we extracted the cancer site, TV type, imaging modality, prompt modality, segmentation architecture, and quantitative performance metrics. DSC was extracted as the primary measure of volumetric overlap between
the predicted and reference contours; HD95 was used as a measure of
boundary accuracy, capturing local surface deviations not reflected by
DSC alone; and sensitivity and precision were extracted to indicate
whether the model tends to undercontour (low sensitivity) or overcontour
(low precision). We also extracted evaluation by a radiation oncologist,
contouring time, and dosimetric outcomes where reported.

Studies were grouped according to prompt modality, and their performance was evaluated against image-only models where reported. Because the included studies differed substantially in cancer sites, datasets, prompting approaches, and evaluation methods, the findings were summarized descriptively rather than by meta-analysis.

%% file: sections/results.tex
\section{Results}
\label{section 4}

\subsection{Study Selection}
\begin{figure*}[!t]
\centering
\includegraphics[width=0.55\textwidth]{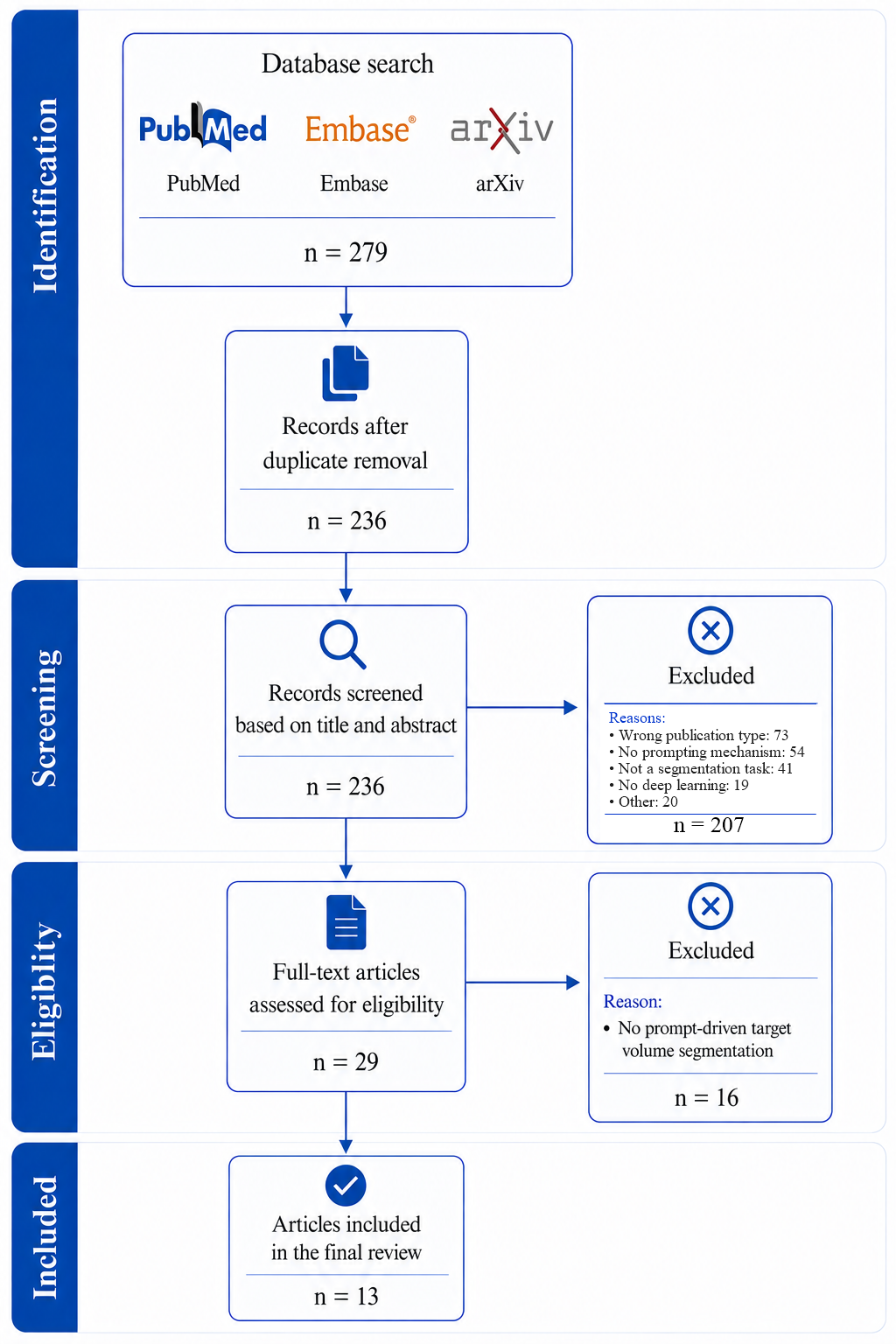}
\caption{PRISMA flow diagram showing the study selection process, from
the 279 records identified through database searches to the thirteen
studies included in the final review. Wrong publication type includes
conference papers and review articles.}

\label{fig:prisma}
\end{figure*}

The database search identified 279 records. After removing 43
duplicates, 236 records underwent title and abstract screening, of
which 29 were assessed for full-text eligibility. The most common
reasons for exclusion at this stage were the article being a
conference paper or review article, the absence of a prompting
mechanism, the study not addressing a segmentation task, and the
absence of a deep learning method. Sixteen studies were subsequently
excluded at full-text review, primarily for the absence of a
prompt-driven segmentation approach for TV delineation, leaving
thirteen studies in the final review. Figure~\ref{fig:prisma} summarizes the study selection process,
including the full breakdown of exclusion reasons.

\subsection{Characteristics of Included Studies}

The thirteen included studies varied in cancer site, TV type,
imaging modality, and prompt modality (Table~\ref{tab:studies}). They
covered six cancer sites: head and neck, breast, prostate, lung, cervix,
and rectum. Seven studies focused on GTV delineation, five on CTV
delineation, and one on both TVs. CT was the predominant imaging
modality, used either alone or in combination with MRI or PET.

In terms of prompt modality, clinical text prompting was the most
common, used in seven studies, followed by geometric prompting in six
studies and interactive prompting in three studies. Three studies
combined multiple prompting strategies: two combined clinical text and
geometric prompts~\citep{peng2026,hooshangnejad2024}, while one combined
geometric and interactive prompts~\citep{nninteractive2025}. Beyond
these characteristics, six studies reported clinical evaluations by radiation oncologists
~\citep{oh2024,khor2025,kim2025,peng2026,blocker2026,nninteractive2025}.

\begin{table*}[!ht]
\centering
\caption{Characteristics of the thirteen included studies. TV: tumor
target volume; GTV: gross tumor volume; CTV: clinical target volume;
DSC: Dice Similarity Coefficient; HD95: 95th percentile Hausdorff
Distance; MSD: mean surface distance; CMD: center-of-mass distance;
IOU: intersection over union; CTime: contouring time; Multi-site: study
evaluated across more than 2 cancer sites.}
\label{tab:studies}
\scriptsize
\renewcommand{\arraystretch}{1.5}
\setlength{\tabcolsep}{8pt}
\begin{tabular}{p{2.5cm} p{1.8cm} p{1.0cm} p{1.7cm} p{2.0cm} p{1.8cm} p{1.2cm}}
\arrayrulecolor{black}\toprule
\textbf{Study} & \shortstack[l]{\textbf{Cancer}\\\textbf{Site}} & \textbf{TV} & \shortstack[l]{\textbf{Imaging}\\\textbf{Modality}} & \shortstack[l]{\textbf{Prompt}\\\textbf{Modality}} & \shortstack[l]{\textbf{Evaluation}\\\textbf{Metrics}} & \shortstack[l]{\textbf{Clinical}\\\textbf{Evaluation}} \\
\arrayrulecolor{black}\midrule
\citet{zhu2025}                    & Head \& Neck     & GTV      & CT, MRI       & Geometric             & DSC, HD95             & No      \\ \arrayrulecolor{gray!40}\hline
\citet{blocker2026}        & Multi-site       & GTV      & MRI           & Geometric             & DSC, HD95             & Yes     \\ \arrayrulecolor{gray!40}\hline
\citet{khor2025}                  & Head \& Neck     & CTV      & CT            & Geometric             & DSC, HD95,\newline IOU        & Yes     \\
\arrayrulecolor{gray!40}\hline
\citet{nninteractive2025}      & Multi-site       & GTV      & CT, MRI       & Geometric\newline Interactive & DSC, CTime    & Yes     \\ \arrayrulecolor{gray!40}\hline
\citet{oh2024}                      & Breast,\newline Prostate & CTV      & CT            & Text         & DSC, HD95             & Yes     \\ \arrayrulecolor{gray!40}\hline
\citet{rajendran2024}       & Prostate,\newline Head \& Neck    & GTV,\newline CTV & MRI, CT       & Text         & DSC, IOU,\newline HD95        & No      \\ \arrayrulecolor{gray!40}\hline
\citet{kim2025}                    & Breast           & CTV      & CT            & Text         & DSC, HD95             & Yes     \\ \arrayrulecolor{gray!40}\hline
\citet{rajendran2025}       & Head \& Neck     & GTV      & CT            & Text         & DSC, IOU,\newline HD95        & No      \\ \arrayrulecolor{gray!40}\hline
\citet{wang2025}                  & Cervix           & CTV      & CT            & Text         & DSC, HD95,\newline MSD, CMD   & No      \\ \arrayrulecolor{gray!40}\hline
\citet{hooshangnejad2024}& Lung             & GTV      & CT            & Text\newline Geometric & DSC, HD95            & No      \\ \arrayrulecolor{gray!40}\hline
\citet{peng2026}                  & Rectum           & CTV      & CT            & Text\newline Geometric & DSC, HD95,\newline MSD, CTime & Yes   \\ \arrayrulecolor{gray!40}\hline
\citet{wei2023}                    & Head \& Neck     & GTV      & CT, PET,\newline MRI  & Interactive  & DSC, HD95,\newline MSD & No    \\ \arrayrulecolor{gray!40}\hline
\citet{saukkoriipi2025}    & Head \& Neck     & GTV      & CT, PET       & Interactive           & DSC, HD95             & No      \\
\arrayrulecolor{black}\bottomrule
\end{tabular}
\end{table*}

Table~\ref{tab:performance} shows the DSC of all thirteen prompt-driven
models alongside their image-only baselines.

\begin{table*}[!t]
\centering
\caption{DSC of prompt-driven models compared with their image-only baselines across the 13 included studies. N/A indicates that no internal image-only baseline was reported in the study.}
\label{tab:performance}
\small
\renewcommand{\arraystretch}{1.4}
\setlength{\tabcolsep}{8pt}
\arrayrulecolor{black}
\begin{tabular}{llcc}
\toprule
\textbf{Study} & \textbf{Cancer Site} & \multicolumn{2}{c}{\textbf{DSC}} \\
\cmidrule(lr){3-4}
 & & \textbf{Prompt model} & \textbf{Image-only baseline} \\
\midrule
\citet{zhu2025}                     & Head \& Neck     & 0.84  & 0.80  \\ \arrayrulecolor{gray!40}\hline
\citet{blocker2026}         & Multi-site       & 0.85  & N/A   \\ \arrayrulecolor{gray!40}\hline
\citet{khor2025}                   & Head \& Neck     & 0.80  & 0.75  \\ \arrayrulecolor{gray!40}\hline
\citet{oh2024}                       & Breast, Prostate & 0.83  & 0.81  \\ \arrayrulecolor{gray!40}\hline
\citet{rajendran2024}         & Prostate, Head \& Neck & 0.81  & 0.72  \\ \arrayrulecolor{gray!40}\hline
\citet{kim2025}                     & Breast           & 0.84  & 0.80  \\ \arrayrulecolor{gray!40}\hline
\citet{rajendran2025}         & Head \& Neck     & 0.76  & 0.74  \\ \arrayrulecolor{gray!40}\hline
\citet{wang2025}                   & Cervix           & 0.80  & 0.79  \\ \arrayrulecolor{gray!40}\hline
\citet{hooshangnejad2024} & Lung             & 0.67  & N/A   \\ \arrayrulecolor{gray!40}\hline
\citet{peng2026}                   & Rectum           & 0.93  & 0.91  \\ \arrayrulecolor{gray!40}\hline
\citet{wei2023}                     & Head \& Neck     & 0.82  & 0.65  \\ \arrayrulecolor{gray!40}\hline
\citet{saukkoriipi2025}     & Head \& Neck     & 0.86  & 0.72  \\ \arrayrulecolor{gray!40}\hline
\citet{nninteractive2025}       & Multi-site       & 0.79  & N/A   \\
\arrayrulecolor{black}\bottomrule
\end{tabular}
\end{table*}

\subsection{Geometric Prompting}
\label{sec:geometric}

Four studies investigated geometric prompting for RT TV delineation, using different forms of spatial guidance to indicate the TV location.

Blocker et al.\ compared SAM2, MedSAM2, nnInteractive, and ContourTracker 3D for GTV delineation on daily MRI scans acquired with MRI-guided linear accelerators~\citep{blocker2026}. Six geometric prompting strategies, ranging from a single center point to three orthogonal binary masks, were simulated from the reference GTV contours. Across 537 patients from six institutions, nnInteractive achieved the highest median DSC, increasing from 0.75 with minimal prompting to 0.85 using the 3-mask prompt, compared with 0.70 for MedSAM2 and 0.54 for SAM2. Performance consistently improved as more spatial information was provided through the prompts.

Khor et al.\ developed RT-SAM for nasopharyngeal carcinoma (NPC) CTV delineation on CT by adapting MedSAM-2 to incorporate automatically generated geometric prompts~\citep{khor2025}. A pretrained U-Net first produces an initial CTV prediction, from which it generates masks, bounding boxes, and point prompts. These prompts are then passed to MedSAM-2 through a visual-prompt fusion attention module. Their framework achieved a mean DSC of 0.80 using five-fold cross-validation on 468 cases, outperforming nnU-Net (DSC 0.75) and Swin-UNETR (0.76). Clinical evaluation by eight radiation oncologists found the generated contours to be clinically acceptable in more than 97\% of cases and generally preferred them over manual delineations.

Isensee et al.\ introduced nnInteractive, a 3D promptable segmentation framework built on nnU-Net and trained on more than 120 datasets~\citep{nninteractive2025}. The framework supports multiple geometric prompt types, including points, bounding boxes, scribbles, and lasso annotations, enabling efficient 3D tumor segmentation from intuitive 2D user inputs. In a radiological user study involving twelve tumor lesions, their framework achieved a median DSC of 0.79 while reducing annotation time by 72\%.

Unlike the preceding studies, Zhu et al.\ did not use explicit geometric prompts such as points, bounding boxes, or scribbles. Instead, they proposed 3DGS-PDM, a diffusion-based framework for NPC GTV segmentation on CT and MRI that generates 3D Gaussian representations from the input images using a Gaussian Initialization Module~\citep{zhu2025}. These Gaussian representations then guide the diffusion model during segmentation. Their framework achieved a mean DSC of 0.84 on 120 test patients from four institutions, outperforming five image-only models.

\subsection{Clinical Text Prompting}
\label{sec:text}
Seven studies investigated the use of clinical text to guide TV delineation. The text prompts were intended to provide information that was not directly available from the medical images alone and were derived from different sources. These sources include electronic medical records, radiology reports, contouring guidelines, and structured clinical information. 

Oh et al.\ first demonstrated the use of patient-specific clinical information to guide RT TV by developing LLMSeg for breast and prostate CTV segmentation~\citep{oh2024}. Structured clinical text, including TNM stage, surgery type, and laterality, extracted from electronic medical records, was encoded and integrated with image features through cross-attention. Their framework achieved a DSC of 0.83 on 307 internal breast cancer patients and maintained similar performance on two external cohorts, achieving DSC values of 0.82 and 0.84 on 206 and 204 patients, respectively. In contrast, the image-only model declined from a DSC of 0.81 on the internal dataset to 0.73 and 0.44 on the external cohorts. Radiation oncologists also rated the LLMSeg contours higher than those generated by the image-only model.

Rajendran et al.\ proposed MedFormer for GTV delineation in prostate and oropharyngeal cancers, using clinical notes rather than structured clinical variables to guide segmentation~\citep{rajendran2024}. Here, GPT-4 was used to extract tumor-related information from the clinical notes, which was then encoded with PubMedBERT and fused with image features through a visual-language attention module. On a prostate MRI dataset of 20 test patients, incorporating clinical text improved the GTV DSC from 0.72 to 0.81. On a CT dataset of 60 patients with oropharyngeal cancer, MedFormer achieved a DSC of 0.77, outperforming the corresponding image-only model.

Rajendran et al.\ subsequently extended this approach to GTV delineation in head-and-neck cancer, proposing RadFormer and evaluated it on the public RADCURE CT dataset of 2,985 patients~\citep{rajendran2025}. Clinical information extracted from patient records was encoded and fused with image features through the same visual-language attention mechanism. Their framework achieved a mean DSC of 0.76 and an HD95 of 7.82 mm on 598 test patients, outperforming nnU-Net (0.74), Swin-UNETR (0.69), and 3D-UNETR (0.66). Removing the language module reduced the DSC from 0.76 to 0.70, demonstrating the contribution of the clinical text to the final segmentation.

Building on LLMSeg~\citep{oh2024}, Kim et al.\ proposed RO-LMM for breast CTV delineation on CT, extending the approach from structured clinical variables to summarized clinical reports~\citep{kim2025}. A fine-tuned LLaMA-2-7B first summarized the clinical reports and generated an RT strategy, which was then encoded and fused with image features to guide the segmentation. Their framework achieved DSC scores of 0.84 and 0.76 on internal and external cohorts of 593 and 81 patients, respectively, outperforming the corresponding image-only model, which achieved DSC values of 0.80 and 0.69. Unlike MedFormer, RadFormer, and LLMSeg, which used lesion-level localization information, RO-LMM first used clinical reports to generate an RT strategy that then guided segmentation, representing a fundamentally different use of clinical text

Unlike the preceding studies, Wang et al.\ derived their text prompts from contouring guidelines rather than patient records, proposing a framework for HR-CTV and OAR delineation in cervical cancer brachytherapy on CT~\citep{wang2025}. ChatGPT was used to extract task-specific instructions from guidelines published by the American Brachytherapy Society, ESTRO, and ICRU. These instructions were encoded with CLIP and fused with image features to guide segmentation. Their framework achieved a DSC of 0.80 for HR-CTV and 0.91 for the bladder using four-fold cross-validation on 32 patients. An ablation study comparing the full framework against a version without text prompt guidance found no statistically significant difference for HR-CTV specifically (0.80 vs 0.79, $p = 0.743$), however, an improvement was observed across all organs combined (mean DSC 0.79 vs 0.78, $p = 0.014$).

Hooshangnejad et al.\ used clinical text to improve tumor localization rather than directly guide segmentation, proposing EXACT-Net for NSCLC GTV detection and segmentation on CT~\citep{hooshangnejad2024}. A Retina-UNet3D first generated candidate tumor regions, after which GPT-3.5 Turbo extracted the tumor lobe location from the clinical reports to remove false-positive candidates before the final UNet3D segmentation. Their framework achieved a DSC of 0.67 on 201 CT scans from the LIDC-IDRI dataset. In a feasibility study involving ten NSCLC patients, incorporating clinical reports increased the correct tumor identification rate from 20\% to 70\%, representing a 250\% improvement over the image-only approach.

Peng et al.\ combined patient-specific clinical information with guideline-derived anatomical specifications to guide CTV delineation in rectal cancer on CT, proposing TG-SegNet~\citep{peng2026}. Structured text prompts containing patient demographics, TNM stage, and the required CTV subsites were encoded with PubMedBERT and integrated with image features and point prompts through cross-attention. Their framework achieved a DSC of 0.93 on 115 test patients, outperforming nnU-Net by approximately 2.1 percentage points. Radiation oncologists rated the generated contours higher for TV coverage, guideline adherence, and clinical acceptability, while the median correction time was reduced to 3.4 minutes, representing an 82\% reduction compared with manual delineation.

\subsection{Interactive Prompting}
\label{sec:interactive}

Only two studies investigated interactive prompting for TV delineation.

Wei et al.\ proposed an inference-time adaptation framework for head and neck (HNC) GTV segmentation, allowing the model to be updated using clinician-provided annotations during inference~\citep{wei2023}. The model first generated an initial segmentation, after which the clinician annotated one or more 2D slices to refine the prediction. Their framework improved the DSC from 0.65 at baseline to 0.82 after three craniocaudally sampled slices across two annotation rounds on 51 test patients, outperforming nnU-Net.

Saukkoriipi et al.\ proposed a 2-stage Interactive Click Refinement (2S-ICR) framework for primary GTV segmentation in oropharyngeal cancer~\citep{saukkoriipi2025}. An initial segmentation generated from CT and PET images was subsequently refined using clinician click prompts. Their framework achieved a DSC of 0.72 without interaction, which increased to 0.86 after ten user interactions on the HECKTOR 2021 dataset and an external cohort from MD Anderson Cancer Center, outperforming DeepGrow and DeepEdit.

%% file: sections/discussion.tex
\section{Discussion}
\label{sec:discussion}
Across the thirteen included studies, TV segmentation models that
incorporated a prompt, whether geometric, clinical text, or
interactive, generally outperformed models trained on imaging data
alone. This supports the central argument of this review: TV
delineation is not a purely visual task, and models that ignore the
clinical reasoning behind a contour are missing information that
imaging alone does not routinely provide, such as tumor biology,
clinical staging, and institutional contouring guidelines. However, the three prompting modalities have largely developed in isolation. Evaluation has focused mostly on geometric overlap, and few studies have examined the effects of incorrect or incomplete prompts. Realizing the clinical potential of PDMs will require closing these gaps.

\subsection{Differences Across Prompt Modalities}
\label{sec:disc_comparison}

Clinical text prompting has been applied across a broader range of
cancer sites than geometric or interactive prompting (Table~\ref{tab:studies}),
both of which remain concentrated in head and neck cancer. This is the only site where all three
strategies have been tested, but never within the same patient
cohort. Without a shared benchmark, it is not possible to determine
which prompting strategy performs best for a given cancer site.

Beyond cancer site, the choice of prompting strategy also appears
linked to the type of TV to which it has most often been applied, and
this pattern follows directly from what each strategy can and cannot provide. Geometric prompts point to a location, so they have mainly guided GTV delineation, where the tumor is visible in the image. On the other hand, clinical text prompts carry information complementary to the image, such as staging and contouring guidelines. Hence, these prompts have been used more often for CTV delineation, where the boundary depends on the clinical context
rather than on what can be seen. One geometric prompting study is an exception that illustrates the point: rather than using manually drawn points or boxes, its prompts were generated automatically from a model's prior prediction, effectively encoding some sense of the expected CTV shape before delineation even began~\citep{khor2025}. Interactive prompting has so far only been applied to GTV delineation, where a clinician corrects the model's initial guess by pointing out where it went wrong rather than supplying the clinical context it was missing.

Clinical text prompting itself has not settled on a single approach
either. Most studies encoded structured clinical variables, such as
TNM stage, tumor location, laterality, and lymph node involvement,
directly or after extracting them from patient records with an
LLM~\citep{hooshangnejad2024,oh2024,peng2026,rajendran2024}. The
remaining studies instead used an LLM to generate a free-form clinical
description from patient records or contouring
guidelines~\citep{kim2025,rajendran2025,wang2025}. No study has
compared these two approaches, so it remains
unclear whether representing clinical knowledge as structured
variables or as LLM-generated text is more effective.

A more flexible form of clinical text prompting is free-text prompting,
in which a clinician describes the target directly in plain language
rather than having the model extract it from a report. This has not
been attempted in RT. One general-purpose segmentation model not built specifically for
RT has used this approach, and despite being trained on a
very large and varied dataset, achieved lower accuracy than
the domain-specific models in this review~\citep{rokuss2025}. This
suggests that the flexibility of free-text prompting may come at the
cost of accuracy without training tailored to the specific clinical
task.

These findings show that prompt-driven TV segmentation has developed
as three separate, largely uncompared lines of research, distinguished
less by the underlying segmentation architecture than by the type of
prompt each relies on and how that prompt is represented.

\subsection{Evaluation Gaps}
\label{sec:disc_evaluation}

All thirteen included studies reported DSC, and twelve also reported
HD95. Geometric overlap is therefore the primary basis for evaluating PDMs. DSC does not distinguish under-contouring from
over-contouring, since a given DSC value can result from either error
type or a combination of both~\citep{reinke2024,maierhein2024}. The sensitivity and precision make this distinction: reduced sensitivity indicates missed tumor tissue and reduced precision indicates the inclusion of healthy
tissue, but neither metric was reported by any of the thirteen
studies.

A geometric score alone may not indicate whether a clinician would
accept the predicted contour. DSC is computed relative to a single reference contour. However,
radiation oncologists often delineate the same TV differently~\citep{guzene2023,jamema2016}. A close match to one reference may therefore not reflect how another radiation oncologist would delineate it. Only six of the thirteen studies assessed clinical acceptability by having a radiation oncologist evaluate the generated contours. Contouring time was reported in only two of the thirteen studies, despite being a common clinical motivation for automating delineation; the remaining studies do not indicate whether their models would reduce clinician workload in practice.

No study assessed dosimetric outcomes. The relevant question is not
only whether an automatically generated contour differs geometrically
from the reference, but whether that difference would produce a
clinically relevant change in the resulting dose distribution relative
to a plan based on the reference contour. Image-based metrics
correlate with segmentation accuracy but do not reliably predict the dose impact ~\citep{rogowski2026}. As with geometric
accuracy, the aim should extend to dosimetric equivalence: treatment
plans generated from prompt-driven contours should be clinically
indistinguishable from those generated using manual contours. The
studies included in this review established the feasibility of
prompt-driven segmentation; the logical next step is to determine whether these geometric
improvements correspond to dosimetric improvements.

The thirteen studies also varied in datasets, patient populations, and
evaluation protocols. This variation was most pronounced for clinical
text prompting, where studies differed in the source of clinical
information, the text encoder used, and the clinical variables
included in the prompt. Because these factors were not varied
systematically, differences in reported performance are difficult to
attribute to any single factor. Standardized evaluation
frameworks~\citep{poel2025}, together with consistent reporting of
prompt construction and clinical variable selection, would improve
reproducibility and comparability across future studies.

\subsection{Robustness to Imperfect Prompts}
\label{sec:disc_robustness}

Few studies examined how PDMs respond to missing or inaccurate prompts, and their conclusions diverged sharply,
ranging from a severe drop in performance to near-complete resilience ~\citep{oh2024,kim2025,peng2026}.
This divergence suggests that robustness is not an inherent property
of using a prompt; instead, it may depend on design choices, such as how a
model is trained to handle missing or noisy input, which most studies
have not investigated. For geometric and interactive prompting, this question was largely
avoided. One study trained its model on deliberately imprecise prompts
but did not evaluate the effect of that imprecision ~\citep{nninteractive2025}. Others explicitly acknowledge that
their simulated interactions never included an incorrect correction,
leaving realistic clinician behavior
untested~\citep{wei2023,saukkoriipi2025}.

This points to a gap broader than robustness alone: robust prompt 
representation and PDM training should be investigated in future research.  Establishing this is a  prerequisite for the clinical introduction of prompt-driven segmentation.

\subsection{Clinical Translation and Future Directions}
\label{sec:disc_future}

The studies included in this review demonstrate that clinical
information not visible on the image can be incorporated into deep
learning models for TV segmentation, generally improving DSC over
image-only baselines (Table~\ref{tab:performance}). However, all
thirteen studies were retrospective and evaluated on single or limited
multi-institutional datasets. No study has reported prospective validation across
institutions, imaging protocols, and clinical workflows for any prompting modality. Generalizability across
different RT centers is therefore the next step towards clinical
translation. Federated learning~\citep{rieke2020} and large
data-sharing initiatives such as MOMENTUM~\citep{demol2020} may
facilitate multi-institutional model development while avoiding or
reducing the need to centralize patient data.

For clinical text prompting, prompt design itself remains an open
research question. The included studies used different sources of
clinical information, different combinations of clinical variables, and
different LLMs to encode them; yet none systematically compared these
choices. Addressing this will require systematic studies of prompt
engineering to establish how clinical information should be selected,
represented, and integrated for different cancer sites and TVs.

Clinical deployment requires more than accurate segmentation. Current
clinical text prompting methods rely on manually prepared text inputs
and operate separately from the RT workflow. In routine practice, the
required clinical information is distributed across electronic health
records, radiology reports, and treatment planning systems. Similarly,
interactive prompting methods must be integrated into RT planning
software while maintaining an efficient user interface and complying
with regulatory requirements, including the EU AI Act~\citep{euaiact2024}.

%% file: sections/conclusion.tex
\section{Conclusion}
\label{sec:conclusion}
We have performed a systematic evaluation of PDMs for TV segmentation.
Models that used a prompt mostly performed better than image-only
models, confirming that clinical context adds real value beyond imaging alone. However, moving from research to clinical
practice will require prospective, multi-institutional validation; a
systematic study of how clinical information should be selected,
represented, and used, and evaluation that extends beyond geometric
accuracy to dosimetric and clinical outcomes.

%% file: sections/declarations.tex
\section*{Declaration of competing interest}
The authors declare that they have no known competing financial interests or personal relationships that could have appeared to influence the work reported in this paper.

\section*{Data availability}
This article is a systematic review. No original data were collected or deposited. The search strategies used to identify eligible studies are provided in Appendix A.

\section*{Funding}
This research received no specific grant from any funding agency in the public, commercial, or not-for-profit sectors.

%% file: sections/appendix.tex
\appendix
\section*{Appendix A}

\subsection*{Search Strategies}
\label{sec:search_strategies}

The following search strategies were used to identify eligible studies. The search was conducted on April 8, 2026, covering the period January 1, 2022 to April 8, 2026.

\vspace{6pt}
\noindent\textbf{Table A.1.} Search strings used across the three databases.
\vspace{4pt}

\begin{center}
\small
\renewcommand{\arraystretch}{1.8}
\begin{tabular}{>{\centering\arraybackslash}m{1.6cm} m{13cm}}
\toprule
\multicolumn{1}{c}{\textbf{Database}} & \multicolumn{1}{c}{\textbf{Search string}} \\
\midrule
PubMed &
\ttfamily\scriptsize
(``radiotherapy''[ti/ab] OR ``radiation therapy''[ti/ab] OR ``radiation oncology''[ti/ab]) AND (``target volume''[ti/ab] OR ``clinical target volume''[ti/ab] OR ``CTV''[ti/ab] OR ``gross tumor volume''[ti/ab] OR ``GTV''[ti/ab]) AND (``segmentation''[ti/ab] OR ``delineation''[ti/ab] OR ``contouring''[ti/ab]) AND (``foundation model''[ti/ab] OR ``multimodal''[ti/ab] OR ``vision-language''[ti/ab] OR ``large language model''[ti/ab] OR ``LLM''[ti/ab] OR ``prompt''[ti/ab] OR ``SAM''[ti/ab] OR ``segment anything''[ti/ab]) AND ``2022/01/01''[Date] : ``2026/04/08''[Date] \\[4pt]
\midrule
Embase &
\ttfamily\scriptsize
(`radiotherapy':ti,ab OR `radiation therapy':ti,ab OR `radiation oncology':ti,ab) AND (`target volume':ti,ab OR `clinical target volume':ti,ab OR `CTV':ti,ab OR `gross tumor volume':ti,ab OR `GTV':ti,ab) AND (`segmentation':ti,ab OR `delineation':ti,ab OR `contouring':ti,ab) AND (`foundation model':ti,ab OR `multimodal':ti,ab OR `vision-language':ti,ab OR `large language model':ti,ab OR `LLM':ti,ab OR `prompt':ti,ab OR `SAM':ti,ab OR `segment anything':ti,ab) AND [2022--2026]/py \\[4pt]
\midrule
arXiv &
\ttfamily\scriptsize
(radiotherapy OR ``radiation therapy'' OR ``radiation oncology'') AND (``target volume'' OR ``clinical target volume'' OR CTV OR ``gross tumor volume'' OR GTV) AND (segmentation OR delineation OR contouring) AND (prompt OR ``large language model'' OR multimodal OR ``vision-language'' OR ``foundation model'' OR ``segment anything'' OR SAM OR LLM); \\
\bottomrule
\end{tabular}
\end{center}
\normalsize

%% file: references.bib
@article{zhu2024global,
  title={Global radiotherapy demands and corresponding radiotherapy-professional workforce requirements in 2022 and predicted to 2050: a population-based study},
  author={Zhu, Hongcheng and Chua, Melvin Lee Kiang and Chitapanarux, Imjai and Kaidar-Person, Orit and Mwaba, Catherine and Alghamdi, Majed and Mignola, Andr{\'e}s Rodr{\'\i}guez and Amrogowicz, Natalia and Yazici, Gozde and Bourhaleb, Zouhour and others},
  journal={The Lancet Global Health},
  volume={12},
  number={12},
  pages={e1945--e1953},
  year={2024},
  publisher={Elsevier},
  doi = {10.1016/s2214-109x(24)00355-3}
}

@article{chen2025,
  author  = {Chen, M. and Peng, Y. and Chen, R. and others},
  title   = {Interobserver variability in organ delineation on radiotherapy treatment planning for nasopharyngeal carcinoma: a dosimetric and prognostic analysis},
  journal = {Front Oncol},
  year    = {2025},
  volume  = {15},
  pages   = {1510568},
  doi = {10.3389/fonc.2025.1510568}
}

@article{peng2018,
  author  = {Peng, Y. and Liu, Y. and Shen, G. and others},
  title   = {Interobserver variations in the delineation of target volumes and organs at risk and their impact on dose distribution in IMRT for nasopharyngeal carcinoma},
  journal = {Oral Oncol},
  year    = {2018},
  volume  = {82},
  pages   = {1--7},
  doi = {10.1016/j.oraloncology.2018.04.025}
}

@article{lucido2025,
  author  = {Lucido, J. J. and Hunt, A. and Gu, B. and others},
  title   = {Multicentre evaluation of deep learning {CT} autosegmentation of the head and neck region for radiotherapy},
  journal = {npj Digit Med},
  year    = {2025},
  volume  = {8},
  pages   = {52},
  doi = {10.1038/s41746-025-01624-z}
}

@inproceedings{kirillov2023,
  author    = {Kirillov, A. and Mintun, E. and Ravi, N. and others},
  title     = {Segment anything},
  booktitle = {Proc IEEE/CVF Int Conf Comput Vis (ICCV)},
  year      = {2023},
  pages     = {4015--4026},
  doi = {10.1109/iccv51070.2023.00371}
}

@article{ma2024,
  author  = {Ma, J. and He, Y. and Li, F. and others},
  title   = {Segment anything in medical images},
  journal = {Nat Commun},
  year    = {2024},
  volume  = {15},
  pages   = {654},
  doi = {10.1038/s41467-024-44824-z}
}

@article{mazurowski2023,
  author  = {Mazurowski, M. A. and Dong, H. and Gu, H. and others},
  title   = {Segment Anything Model for medical image analysis: an experimental study},
  journal = {Med Image Anal},
  year    = {2023},
  volume  = {89},
  pages   = {102918},
  doi = {10.1016/j.media.2023.102918}
}

@article{cheng2023,
  author  = {Cheng, J. and Ye, J. and Deng, Z. and others},
  title   = {{SAM-Med2D}},
  journal = {arXiv:2308.16184},
  year    = {2023},
  url = {https://arxiv.org/abs/2308.16184}
}

@article{huang2024,
  author  = {Huang, Y. and Yang, X. and Liu, L. and others},
  title   = {Segment anything model for medical images?},
  journal = {Med Image Anal},
  year    = {2024},
  volume  = {92},
  pages   = {103061},
  doi = {10.1016/j.media.2023.103061}
}

@inproceedings{wang2023,
  title={SAM-Med3D: Towards General-Purpose Segmentation Models for Volumetric Medical Images},
  author={Wang, Haoyu and Guo, Sizheng and Ye, Jin and Deng, Zhongyi and Cheng, Junlong and Li, Tian-Xin and Chen, Jianpin and Su, Yan-Cheng and Huang, Ziyan and Shen, Yiqing and others},
  booktitle={Proceedings of the IEEE/CVF International Conference on Computer Vision (ICCV) Workshops},
  year={2023},
  doi = {10.1007/978-3-031-91721-9_4}
}

@article{oh2024,
  title={LLM-driven multimodal target volume contouring in radiation oncology},
  author={Oh, Yujin and Park, Sangjoon and Byun, Hwa Kyung and Cho, Yeona and Lee, Ik Jae and Kim, Jin Sung and Ye, Jong Chul},
  journal={Nature Communications},
  volume={15},
  number={1},
  pages={9186},
  year={2024},
  publisher={Nature Publishing Group UK London},
  doi = {10.1038/s41467-024-53387-y}
}

@article{rajendran2025,
  title={Large language model-augmented learning for auto-delineation of treatment targets in head-and-neck cancer radiotherapy},
  author={Rajendran, Praveenbalaji and Yang, Yong and Niedermayr, Thomas R and Gensheimer, Michael and Beadle, Beth and Le, Quynh-Thu and Xing, Lei and Dai, Xianjin},
  journal={Radiotherapy and Oncology},
  volume={205},
  pages={110740},
  year={2025},
  publisher={Elsevier},
  doi = {10.2139/ssrn.4956514}
}

@article{nninteractive2025,
  author  = {Isensee, F. and Rokuss, M. and Kr\"{a}mer, M. and others},
  title   = {nnInteractive: Redefining {3D} promptable segmentation},
  journal = {arXiv:2503.08373},
  year    = {2025},
  url = {https://arxiv.org/abs/2503.08373}
}

@article{poel2025,
  author  = {Poel, R. and R\"{u}fenacht, E. and Scheib, S. and others},
  title   = {A comprehensive multifaceted technical evaluation framework for implementation of auto-segmentation models in radiotherapy ({COMMUTE})},
  journal = {Commun Med},
  year    = {2025},
  volume  = {5},
  pages   = {25},
  doi = {10.1038/s43856-025-01048-6}
}

@techreport{euaiact2024,
  author      = {{European Parliament and Council}},
  title       = {Regulation ({EU}) 2024/1689 of the European Parliament and of the Council laying down harmonised rules on Artificial Intelligence (Artificial Intelligence Act)},
  institution = {Official Journal of the European Union},
  year        = {2024}
}

@article{rieke2020,
  author  = {Rieke, N. and Hancox, J. and Li, W. and others},
  title   = {The future of digital health with federated learning},
  journal = {npj Digit Med},
  year    = {2020},
  volume  = {3},
  pages   = {119},
  doi = {10.1038/s41746-020-00323-1}
}

@article{demol2020,
  author  = {de Mol van Otterloo, S.R. and Christodouleas, J.P. and Blezer, E.L.A. and Akhiat, H. and Brown, K. and Choudhury, A. and Eggert, D. and Erickson, B.A. and Faivre-Finn, C. and Fuller, C.D. and Goldwein, J. and Hafeez, S. and Hall, E. and Harrington, K.J. and van der Heide, U.A. and Huddart, R.A. and Intven, M.P.W. and Kirby, A.M. and Lalondrelle, S. and McCann, C. and Minsky, B.D. and Mook, S. and Nowee, M.E. and Oelfke, U. and Orrling, K. and Sahgal, A. and Sarmiento, J.G. and Schultz, C.J. and Tersteeg, R.J.H.A. and Tijssen, R.H.N. and Tree, A.C. and van Triest, B. and Hall, W.A. and Verkooijen, H.M.},
  title   = {The {MOMENTUM} Study: An International Registry for the Evidence-Based Introduction of {MR}-Guided Adaptive Therapy},
  journal = {Front Oncol},
  year    = {2020},
  volume  = {10},
  pages   = {1328},
  doi     = {10.3389/fonc.2020.01328}
}

@techreport{icru50,
  author = {{International Commission on Radiation Units and Measurements}},
  title = {{ICRU} Report 50: Prescribing, Recording, and Reporting Photon Beam Therapy},
  institution = {ICRU},
  address = {Bethesda, MD},
  year = {1993}
}

@techreport{icru62,
  author = {{International Commission on Radiation Units and Measurements}},
  title = {{ICRU} Report 62: Prescribing, Recording and Reporting Photon Beam Therapy (Supplement to {ICRU} Report 50)},
  institution = {ICRU},
  address = {Bethesda, MD},
  year = {1999}
}

@techreport{icru83,
  author = {{International Commission on Radiation Units and Measurements}},
  title = {{ICRU} Report 83: Prescribing, Recording, and Reporting Intensity-Modulated Photon-Beam Therapy ({IMRT})},
  institution = {ICRU},
  address = {Bethesda, MD},
  year = {2010}
}

@article{guzene2023,
  author = {Guzene, L. and Beddok, A. and Nioche, C. and others},
  title = {Assessing interobserver variability in the delineation of structures in radiation oncology: a systematic review},
  journal = {Int. J. Radiat. Oncol. Biol. Phys.},
  year = {2023},
  volume = {115},
  number = {5},
  pages = {1047--1060},
  doi = {10.1016/j.ijrobp.2022.11.021}
}

@article{jamema2016,
  author = {Segedin, B. and Petric, P.},
  title = {Uncertainties in target volume delineation in radiotherapy: are they relevant and what can we do about them?},
  journal = {Radiol. Oncol.},
  year = {2016},
  volume = {50},
  number = {3},
  pages = {254--262},
  doi = {10.1515/raon-2016-0023}
}

@article{litjens2017,
  author = {Litjens, G. and Kooi, T. and Bejnordi, B. E. and others},
  title = {A survey on deep learning in medical image analysis},
  journal = {Med. Image Anal.},
  year = {2017},
  volume = {42},
  pages = {60--88},
  doi = {10.1016/j.media.2017.07.005}
}

@inproceedings{ronneberger2015,
  author = {Ronneberger, O. and Fischer, P. and Brox, T.},
  title = {{U-Net}: convolutional networks for biomedical image segmentation},
  booktitle = {Proc. MICCAI},
  year = {2015},
  pages = {234--241},
  doi = {10.1007/978-3-319-24574-4_28}
}

@inproceedings{cicek2016,
  author = {{\c{C}}i{\c{c}}ek, {\"O}. and Abdulkadir, A. and Lienkamp, S. S. and others},
  title = {{3D U-Net}: learning dense volumetric segmentation from sparse annotation},
  booktitle = {Proc. MICCAI},
  year = {2016},
  pages = {424--432},
  doi = {10.1007/978-3-319-46723-8_49}
}

@article{isensee2021,
  author = {Isensee, F. and Jaeger, P. F. and Kohl, S. A. A. and others},
  title = {{nnU-Net}: a self-configuring method for deep learning-based biomedical image segmentation},
  journal = {Nat. Methods},
  year = {2021},
  volume = {18},
  pages = {203--211},
  doi = {10.1038/s41592-020-01008-z}
}

@inproceedings{hatamizadeh2022,
  author = {Hatamizadeh, A. and Tang, Y. and Nath, V. and others},
  title = {{UNETR}: transformers for {3D} medical image segmentation},
  booktitle = {Proc. IEEE/CVF Winter Conf. Appl. Comput. Vis.},
  year = {2022},
  pages = {574--584},
  doi = {10.1109/wacv51458.2022.00181}
}

@article{cardenas2019,
  author = {Cardenas, C. E. and Yang, J. and Anderson, B. M. and Court, L. E. and Brock, K. B.},
  title = {Advances in auto-segmentation},
  journal = {Semin. Radiat. Oncol.},
  year = {2019},
  volume = {29},
  number = {3},
  pages = {185--197},
  doi = {10.1016/j.semradonc.2019.02.001}
}

@article{doolan2023,
  author = {Doolan, P. J. and Charalambous, S. and Roussakis, Y. and others},
  title = {A clinical evaluation of the performance of five commercial artificial intelligence contouring systems for radiotherapy},
  journal = {Front. Oncol.},
  year = {2023},
  volume = {13},
  pages = {1213068},
  doi = {10.3389/fonc.2023.1213068}
}

@article{debooij2024,
  author = {De Biase, Alessia and Sijtsema, Nanna Maria and Janssen, Tomas and Hurkmans, Coen and Brouwer, Charlotte and van Ooijen, Peter},
  title = {Clinical adoption of deep learning target auto-segmentation for radiation therapy: challenges, clinical risks, and mitigation strategies},
  journal = {BJR Artif. Intell.},
  year = {2024},
  volume = {1},
  number = {1},
  pages = {ubae015},
  doi = {10.1093/bjrai/ubae015}
}

@article{matoska2024,
  author  = {Matoska, T. and Patel, M. and Liu, H. and Beriwal, S.},
  title   = {Review of deep learning based autosegmentation for clinical target volume: current status and future directions},
  journal = {Adv. Radiat. Oncol.},
  year    = {2024},
  volume  = {9},
  number  = {5},
  pages   = {101470},
  doi     = {10.1016/j.adro.2024.101470}
}

@inproceedings{milletari2016,
  author    = {Milletari, F. and Navab, N. and Ahmadi, S. A.},
  title     = {{V-Net}: fully convolutional neural networks for volumetric medical image segmentation},
  booktitle = {Proc. 4th Int. Conf. 3D Vision (3DV)},
  year      = {2016},
  pages     = {565--571},
  doi       = {10.1109/3dv.2016.79}
}

@inproceedings{oktay2018,
  author    = {Oktay, O. and Schlemper, J. and Le Folgoc, L. and others},
  title     = {Attention {U-Net}: learning where to look for the pancreas},
  booktitle = {Proc. Med. Imaging Deep Learn. (MIDL)},
  year      = {2018},
  url       = {https://openreview.net/forum?id=Skft7cijM}
}

@inproceedings{zhou2018,
  author = {Zhou, Z. and Rahman Siddiquee, M. M. and Tajbakhsh, N. and Liang, J.},
  title = {{UNet++}: a nested {U-Net} architecture for medical image segmentation},
  booktitle = {Deep Learning in Medical Image Analysis and Multimodal Learning for Clinical Decision Support},
  year = {2018},
  pages = {3--11},
  doi = {10.1007/978-3-030-00889-5_1}
}

@article{chen2021,
  author = {Chen, J. and Lu, Y. and Yu, Q. and others},
  title = {{TransUNet}: transformers make strong encoders for medical image segmentation},
  journal = {arXiv:2102.04306},
  year = {2021},
  url = {https://arxiv.org/abs/2102.04306}
}

@inproceedings{tang2022,
  author    = {Tang, Y. and Yang, D. and Li, W. and others},
  title     = {Self-supervised pre-training of {Swin} transformers for {3D} medical image analysis},
  booktitle = {Proc. IEEE/CVF Conf. Comput. Vis. Pattern Recognit. (CVPR)},
  year      = {2022},
  pages     = {20730--20740},
  doi       = {10.1109/cvpr52688.2022.02007}
}

@article{zhou2023,
  author  = {Zhou, H. Y. and Guo, J. and Zhang, Y. and others},
  title   = {{nnFormer}: interleaved transformer for volumetric segmentation},
  journal = {IEEE Trans. Med. Imaging},
  year    = {2023},
  volume  = {42},
  number  = {6},
  pages   = {1735--1745},
  doi     = {10.1109/TMI.2023.3243188}
}

@article{page2021,
  author = {Page, M. J. and McKenzie, J. E. and Bossuyt, P. M. and others},
  title = {The {PRISMA} 2020 statement: an updated guideline for reporting systematic reviews},
  journal = {BMJ},
  year = {2021},
  volume = {372},
  pages = {n71},
  doi = {10.1136/bmj.n71}
}

@misc{rayyan,
  author       = {{Rayyan Systems Inc.}},
  title        = {Rayyan},
  year         = {2016},
  howpublished = {\url{https://www.rayyan.ai}},
  note         = {[Software]}
}

@article{wei2023,
  author  = {Wei, Z. and Ren, J. and Korreman, S. S. and Nijkamp, J.},
  title   = {Towards interactive deep-learning for tumour segmentation in head and neck cancer radiotherapy},
  journal = {Phys Imaging Radiat Oncol},
  year    = {2023},
  volume  = {25},
  pages   = {100408},
  doi     = {10.1016/j.phro.2022.12.005}
}

@article{kim2025,
  author  = {Kim, K. and Oh, Y. and Park, S. and Byun, H. K. and Lee, J. and Kim, J. S. and Ye, J. C.},
  title   = {End-to-end breast cancer radiotherapy planning via {LMMs} with consistency embedding},
  journal = {Med Image Anal},
  year    = {2025},
  volume  = {105},
  pages   = {103646},
  doi     = {10.1016/j.media.2025.103646}
}

@article{blocker2026,
  author  = {Bl{\"o}cker, T. J. and others},
  title   = {{GTV} segmentation in {MRI}-guided radiotherapy with promptable foundation models},
  journal = {Phys Med Biol},
  year    = {2026},
  volume  = {71},
  pages   = {015006}
}

@article{zhu2025,
  author  = {Zhu, J. and Ma, Z. and Ren, G. and Cai, J.},
  title   = {Interpretable and performant multimodal nasopharyngeal carcinoma {GTV} segmentation with clinical priors guided {3D-Gaussian-Prompted} diffusion model ({3DGS-PDM})},
  journal = {Cancers},
  year    = {2025},
  volume  = {17},
  pages   = {3660},
  doi     = {10.3390/cancers17223660}
}

@article{saukkoriipi2025,
  author  = {Saukkoriipi, M. and Sahlsten, J. and Jaskari, J. and others},
  title   = {Interactive {3D} segmentation for primary gross tumor volume in oropharyngeal cancer},
  journal = {Sci Rep},
  year    = {2025},
  volume  = {15},
  pages   = {28589},
  doi     = {10.1038/s41598-025-13601-3}
}

@article{wang2025,
  author  = {Wang, J. and Zhang, J. and Yang, K. and others},
  title   = {Enhancing auto-contouring with large language model in high-dose rate brachytherapy for cervical cancers},
  journal = {Med Phys},
  year    = {2025},
  volume  = {52},
  pages   = {e70034},
  doi     = {10.1002/mp.70034}
}

@article{peng2026,
  author  = {Peng, H. and others},
  title   = {Text-guided automatic segmentation of clinical target volume in rectal cancer radiotherapy},
  journal = {Phys Med Biol},
  year    = {2026},
  volume  = {71},
  pages   = {075009}
}

@article{rajendran2024,
  author  = {Rajendran, P. and Chen, Y. and Qiu, L. and others},
  title   = {Auto-delineation of treatment target volume for radiation therapy using large language model-aided multimodal learning},
  journal = {Int J Radiat Oncol Biol Phys},
  year    = {2025},
  volume  = {121},
  number  = {1},
  pages   = {230--240},
  doi     = {10.1016/j.ijrobp.2024.07.2149}
}

@article{hooshangnejad2024,
  author  = {Hooshangnejad, H. and Feng, X. and Huang, G. and others},
  title   = {{EXACT-Net}: {EHR}-guided lung tumor auto-segmentation for non-small cell lung cancer radiotherapy},
  journal = {arXiv:2402.14099},
  year    = {2024},
  url     = {https://arxiv.org/abs/2402.14099}
}

@article{khor2025,
  author  = {Khor, H. G. and Yang, X. and Sun, Y. and others},
  title   = {{RT-SAM}: visual-prompt fusion and uncertainty enhancement for nasopharyngeal carcinoma radiotherapy target delineation},
  journal = {IEEE J Biomed Health Inform},
  year    = {2025},
  doi     = {10.1109/JBHI.2026.3699979}
}

@article{rokuss2025,
  author  = {Rokuss, M. and Langenberg, M. and Kirchhoff, Y. and others},
  title   = {{VoxTell}: free-text promptable universal {3D} medical image segmentation},
  journal = {arXiv:2511.11450},
  year    = {2025},
  url     = {https://arxiv.org/abs/2511.11450}
}

@inproceedings{devlin2019,
  author    = {Devlin, J. and Chang, M.-W. and Lee, K. and Toutanova, K.},
  title     = {{BERT}: Pre-training of deep bidirectional transformers for language understanding},
  booktitle = {Proc. NAACL-HLT},
  year      = {2019},
  pages     = {4171--4186},
  doi       = {10.18653/v1/N19-1423}
}

@article{openai2023,
  author  = {{OpenAI}},
  title   = {{GPT-4} Technical Report},
  journal = {arXiv:2303.08774},
  year    = {2023},
  url     = {https://arxiv.org/abs/2303.08774}
}

@article{luo2022,
  author  = {Luo, R. and Sun, L. and Xia, Y. and others},
  title   = {{BioGPT}: Generative pre-trained transformer for biomedical text generation and mining},
  journal = {Brief. Bioinform.},
  year    = {2022},
  volume  = {23},
  number  = {6},
  doi     = {10.1093/bib/bbac409}
}

@inproceedings{radford2021,
  author    = {Radford, A. and Kim, J. W. and Hallacy, C. and others},
  title     = {Learning transferable visual models from natural language supervision},
  booktitle = {Proc. Int. Conf. Mach. Learn. (ICML)},
  year      = {2021},
  pages     = {8748--8763},
  url       = {https://arxiv.org/abs/2103.00020}
}

@inproceedings{liu2023llava,
  author    = {Liu, H. and Li, C. and Wu, Q. and Lee, Y. J.},
  title     = {Visual instruction tuning},
  booktitle = {Adv. Neural Inf. Process. Syst.},
  year      = {2023},
  volume    = {36},
  url       = {https://arxiv.org/abs/2304.08485}
}

@inproceedings{li2022blip,
  author    = {Li, J. and Li, D. and Xiong, C. and Hoi, S.},
  title     = {{BLIP}: Bootstrapping language-image pre-training for unified vision-language understanding and generation},
  booktitle = {Proc. Int. Conf. Mach. Learn. (ICML)},
  year      = {2022},
  pages     = {12888--12900},
  url       = {https://arxiv.org/abs/2201.12086}
}

@article{ravi2024,
  author  = {Ravi, N. and Gabeur, V. and Hu, Y.-T. and others},
  title   = {{SAM} 2: Segment anything in images and videos},
  journal = {arXiv:2408.00714},
  year    = {2024},
  url     = {https://arxiv.org/abs/2408.00714}
}

@inproceedings{liu2021swin,
  author    = {Liu, Z. and Lin, Y. and Cao, Y. and others},
  title     = {Swin Transformer: Hierarchical vision transformer using shifted windows},
  booktitle = {Proc. IEEE/CVF Int. Conf. Comput. Vis. (ICCV)},
  year      = {2021},
  pages     = {10012--10022},
  doi       = {10.1109/ICCV48922.2021.00986}
}

@inproceedings{zhou2022coop,
  author    = {Zhou, K. and Yang, J. and Loy, C. C. and Liu, Z.},
  title     = {Learning to prompt for vision-language models},
  booktitle = {Int. J. Comput. Vis.},
  year      = {2022},
  volume    = {130},
  pages     = {2337--2348},
  doi       = {10.1007/s11263-022-01653-1}
}

@article{vollenbrock2019,
  author  = {Vollenbrock, S. E. and Nowee, M. E. and Voncken, F. E. M. and Kotte, A. N. T. J. and Goense, L. and van Rossum, P. S. N. and van Lier, A. L. H. M. W. and Heijmink, S. W. and Bartels-Rutten, A. and Wessels, F. J. and Aleman, B. M. P. and Dewit, L. and Kerkmeijer, L. G. W. and Jansen, E. P. M. and Intven, M. and Lips, I. M. and Meijer, G. J. and Nijkamp, J.},
  title   = {Gross tumor delineation in esophageal cancer on {MRI} compared with {18F-FDG-PET/CT}},
  journal = {Adv. Radiat. Oncol.},
  year    = {2019},
  volume  = {4},
  number  = {4},
  pages   = {596--604},
  doi     = {10.1016/j.adro.2019.04.004}
}

@article{lee2020biobert,
  author  = {Lee, J. and Yoon, W. and Kim, S. and others},
  title   = {{BioBERT}: a pre-trained biomedical language representation model for biomedical text mining},
  journal = {Bioinformatics},
  year    = {2020},
  volume  = {36},
  number  = {4},
  pages   = {1234--1240},
  doi     = {10.1093/bioinformatics/btz682}
}

@article{siegel2024,
  author  = {Siegel, R. L. and Giaquinto, A. N. and Jemal, A.},
  title   = {Cancer statistics, 2024},
  journal = {CA Cancer J. Clin.},
  year    = {2024},
  volume  = {74},
  number  = {1},
  pages   = {12--49},
  doi     = {10.3322/caac.21820}
}

@book{brierley2017,
  author    = {Brierley, J. D. and Gospodarowicz, M. K. and Wittekind, C.},
  title     = {{TNM} Classification of Malignant Tumours},
  edition   = {8th},
  publisher = {Wiley-Blackwell},
  year      = {2017}
}

@article{shen2017,
  author  = {Shen, D. and Wu, G. and Suk, H.-I.},
  title   = {Deep learning in medical image analysis},
  journal = {Annu. Rev. Biomed. Eng.},
  year    = {2017},
  volume  = {19},
  pages   = {221--248},
  doi     = {10.1146/annurev-bioeng-071516-044442}
}

@article{reinke2024,
  author  = {Reinke, Annika and Tizabi, Minu D. and Baumgartner, Michael and Eisenmann, Matthias and Heckmann-N{\"o}tzel, Doreen and Kavur, A. Emre and R{\"a}dsch, Tim and Sudre, Carole H. and Acion, Laura and Antonelli, Michela and Arbel, Tal and Bakas, Spyridon and others},
  title   = {Understanding metric-related pitfalls in image analysis validation},
  journal = {Nat. Methods},
  year    = {2024},
  volume  = {21},
  number  = {2},
  pages   = {182--194},
  doi     = {10.1038/s41592-023-02150-0}
}

@article{maierhein2024,
  author  = {Maier-Hein, Lena and Reinke, Annika and Godau, Patrick and Tizabi, Minu D. and Buettner, Florian and Christodoulou, Evangelia and Glocker, Ben and Isensee, Fabian and Kleesiek, Jens and Kozubek, Michal and Reyes, Mauricio and Riegler, Michael A. and Wiesenfarth, Manuel and Kavur, A. Emre and Sudre, Carole H. and Baumgartner, Michael and Eisenmann, Matthias and Heckmann-N{\"o}tzel, Doreen and R{\"a}dsch, Tim and Acion, Laura and Antonelli, Michela and Arbel, Tal and Bakas, Spyridon and Benis, Arriel and Blaschko, Matthew B. and Cardoso, M. Jorge and Cheplygina, Veronika and Cimini, Beth A. and Collins, Gary S. and Farahani, Keyvan and Ferrer, Luciana and Galdran, Adrian and van Ginneken, Bram and Haase, Robert and Hashimoto, Daniel A. and Hoffman, Michael M. and Huisman, Merel and Jannin, Pierre and Kahn, Charles E. and Kainmueller, Dagmar and Kainz, Bernhard and Karargyris, Alexandros and Karthikesalingam, Alan and Kofler, Florian and Kopp-Schneider, Annette and Kreshuk, Anna and Kurc, Tahsin and Landman, Bennett A. and Litjens, Geert and Madani, Amin and Maier-Hein, Klaus and Martel, Anne L. and Mattson, Peter and Meijering, Erik and Menze, Bjoern and Moons, Karel G. M. and M{\"u}ller, Henning and Nichyporuk, Brennan and Nickel, Felix and Petersen, Jens and Rajpoot, Nasir and Rieke, Nicola and Saez-Rodriguez, Julio and S{\'a}nchez, Clara I. and Shetty, Shravya and van Smeden, Maarten and Summers, Ronald M. and Taha, Abdel A. and Tiulpin, Aleksei and Tsaftaris, Sotirios A. and Van Calster, Ben and Varoquaux, Ga{\"e}l and J{\"a}ger, Paul F.},
  title   = {Metrics reloaded: recommendations for image analysis validation},
  journal = {Nat. Methods},
  year    = {2024},
  volume  = {21},
  number  = {2},
  pages   = {195--212},
  doi     = {10.1038/s41592-023-02151-z}
}

@article{rogowski2026,
  author  = {Rogowski, Viktor and Terpstra, Maarten L. and Wahl, Niklas and Kamp, Florian and van der Bijl, Erik and Galapon, Arthur Jr. and Kurz, Christopher and Xin, Bowen and Sun, Zhengxiang and Min, Hollie and Belous, Gregg and Dowling, Jason and Xia, Yan and Mei, Siyuan and Fan, Fuxin and Longuefosse, Arthur and Sequeiro Gonzalez, Javier and Diaz Benito, Miguel and Garcia Martin, Alvaro and Baldacci, Fabien and Boussot, Valentin and H{\'e}mon, C{\'e}dric and Nunes, Jean-Claude and Dillenseger, Jean-Louis and Zhang, Zhiyuan and Cai, Jinghua and Bing, Han and Zuopeng, Tan and Brioso, Ricardo and Loiacono, Daniele and Landry, Guillaume and Thummerer, Adrian and Maspero, Matteo},
  title   = {Generating synthetic computed tomography for radiotherapy: SynthRAD2025 challenge report},
  journal = {arXiv preprint arXiv:2605.13555},
  year    = {2026},
  url     = {https://arxiv.org/abs/2605.13555}
}

@article{darby2026,
  author  = {Darby, Polly and Kilgour, Emily and Then, Chee Kin and Bromiley, Andrew and McLellan, John and Kiltie, Anne E.},
  title   = {Inter-observer variability in radiotherapy contouring with the use of autocontouring software: A systematic review},
  journal = {Clin Transl Radiat Oncol},
  year    = {2026},
  volume  = {58},
  pages   = {101144},
  doi     = {10.1016/j.ctro.2026.101144}
}
